\documentclass[conference]{IEEEtran}

\usepackage[usenames,dvipsnames]{pstricks}
 \usepackage{epsfig}
 \usepackage{pst-plot} 
 \usepackage{pst-grad} 
 \usepackage{mathrsfs,amsmath}%
\usepackage{float}
\usepackage{amsfonts}%
\usepackage{amssymb}%
\usepackage{graphicx}
\usepackage{cite}

\ifCLASSINFOpdf
\else
\fi
\begin{document}

\title{A Statistical Block Fading Channel Model for Multiuser Massive MIMO System}

\author{\IEEEauthorblockN{Suresh Dahiya}
\IEEEauthorblockA{Indian Institute of Technology, Jodhpur, India\\
Email: pg201282009@iitj.ac.in}
}

\maketitle



\begin{abstract}
This paper presents a statistical block fading channel model for multiuser massive MIMO system. The proposed channel model is evolved from correlation based stochastic channel model (CBSCM) but in addition to the properties of CBSCM, it has capability of capturing channel variations along time or frequency and along space simultaneously. It has a simplified analytical expression, still being able to simulate underlying physical phenomena which otherwise need a complex geometry based stochastic channel model (GBSCM). The channel model is verified with reported measurement data of channel for massive MIMO. Spatial determinism in channel, the basic cause of unfavorable propagation, is modeled into controlling parameters of channel model. Channel model uses only three controlling parameters; one parameter describes variation in channel along resource block (along time or frequency) and remaining two parameters describe spatial variation in channel. Modeling of simultaneous variation along time and space belongs to a very common scenario where mobility of mobile terminal and angular power distribution at base station receiver,	 are key parameters. Additionally, simulation results reveal the hidden advantages of spatial determinism in channel for multiuser massive MIMO.

\end{abstract}
\IEEEpeerreviewmaketitle


\renewcommand{\baselinestretch}{0.8}

\section{Introduction}
Massive multiple input multiple output (MIMO) system has great potential for 5th generation wireless communication technologies \cite{LuLu2014,Larsson2014,Bjornson2015_MMyths,ScaleupMIMO2013,LMS_AChockalingam} where base station (BS) equipped with 100's of antennas serves 10's of mobile terminals (MTs) with single or few antennas. Current standards like LTE, 802.11 (WiFi), 802.16 (WiMAX) etc already have provision of up to 8 antennas \cite{Ngo2013a}. Massive MIMO is front runner among communication technologies for future in context of data rates, spectral efficiency, energy efficiency and cancellation of channel fades \cite{Marzett2010,Ngo2013}. Large scaling of antenna has put several new issues, some of which like channel modeling, antennas space constraint, channel training, limited coherence time at higher mobility and higher frequency, still need an intensified research. \\

Channel modeling for massive MIMO is one of the basic issues raised in \cite{LuLu2014} because it provides an analytical testing framework for the designers of communication schemes and cellular structure. However, channel models for small scale MIMO are in literature for a long time like a correlation based stochastic channel model (CBSCM) for wide band MIMO system, is provided in \cite{Hong2007} and a geometry based stochastic channel model (GBSCM) for multi-link MIMO communication is provided in \cite{Poutanen2012}. But, the aforementioned channel models are not directly suitable for massive MIMO. Theoretical studies for massive MIMO typically consider either independent and identically distributed (i.i.d.) complex Gaussian channel model alone \cite{Marzett2010} or with multiplication of a diagonal matrix describing large scale fading \cite{Ngo2013}.  A need for new channel model for massive MIMO is suggested in \cite{Payami2012} based on measured channel. A non-stationary 3-D wideband twin-cluster model for massive MIMO is provided in \cite{Wu2014} where most of the underlying physical phenomena are included in channel model which can provide a good framework if different propagation parameters like mobility, geometry of clusters etc. can be collected precisely. However, managing all these parameters while analyzing a communication scheme using this channel model, is again a challenge. Moreover, a better insights on a communication scheme can be obtained by using a simplified analytical channel model \cite{Bjornson2015}. However, the channel model used in \cite{Bjornson2015}, has implicit assumption of i.i.d. channel coefficients and it was specific to non line of sight (NLOS) scenario. On the other hand, CBSCMs are not sufficient to model spatial determinism in the channel as well as CBSCMs can simulate correlation along only one dimension. Thus, there is a need for a channel model which can provide a simplified analytical expression for the channel and can simulate temporal, frequential and spatial variation simultaneously. However, even doing so, it must be able to simulate underlying propagation phenomena like the \cite{Wu2014}.\\

This paper presents a statistical analytical channel model for massive MIMO where different underlying physical phenomena are modeled into statistical parameters of channel model while keeping the controlling parameters minimum. This work considers a multiuser massive MIMO system with block fading channel. Block fading assumption for channel, resembles with practical scenario for cellular communication e.g. LTE. The important aspect of this channel model is that the most of the details of channel are folded into three controlling parameters (matrices) making the channel model suitable for analytic evaluation of communication schemes. Parameters of analytic expression of channel model, are well connected to temporal, frequential and spatial variations. This work reveals the hidden aspects of antenna-phase relationship which was not observed in experimental measurement of \cite{Payami2012}, possibly due to longer period per measurement (around half an hour). Simulation results on distribution of eigenvalues, user-correlation and average received power along array dimension, resemble with measured data \cite{Payami2012,Gao2015}. This channel model identifies the spatial determinism property of channel vectors which provides a new insight on propagation aspects. Simulation result on average statistics of channel, again, provides a new insight on unfavorable propagation (cf. \cite{ScaleupMIMO2013}) aspects of channel.\\

Rest of the paper is organized in following sections. Section II describes the system model with detailed explanation of underlying physical phenomena, Section III shows insights on simulated channel with proper explanation, Section IV validates the channel model by comparing parameters of simulated channel model with existing measured channel parameters and Section V draws out conclusion.

\section{System Model}
Considering a practical system like LTE, the time-frequency resources (total of $t_{mx}\times{f_{mx}}$) for a wireless cellular system, under block fading, can be divided into RBs as follows.
\begin{center}
$
\begin{bmatrix}
	(1,1) 			& (1,2) 			& ... & (1,f_{mx})\\
	(2,1) 			& (2,2) 			& ... & (2,f_{mx})\\
	....      	& ....      	& ... & ....\\
	(t_{mx},1) & (t_{mx},2) & ... & (t_{mx},f_{mx})
\end{bmatrix}
$
\end{center}

Let a cellular massive MIMO system has $K$ active mobile terminals (MTs) and one BS with $N$ antennas. If $T$ number of symbols are received at BS in each RB with index $(t,f)$ then, uplink received matrix at BS can be written as.

\begin{center}
$\mathbf{Y}(t,f)=\mathbf{H}(t,f)\mathbf{X}(t,f)+\mathbf{W}(t,f)$ where
\end{center}
$\mathbf{Y}(t,f)$ is $N\times{T}$ received complex matrix in $(t,f)^{th}$ RB.\\
$\mathbf{X}(t,f)$ is $K\times{T}$ transmitted complex matrix in $(t,f)^{th}$ RB such that $\mathbb{E}[x_{ij}(t,f)]=1$.\\
$\mathbf{W}(t,f)$ is $N\times{T}$ complex AWGN matrix having i.i.d. entries in $(t,f)^{th}$ RB with zero mean \& finite variance (measure of SNR).\\
$\mathbf{H}(t,f)$ is $N\times{K}$ complex channel matrix in $(t,f)^{th}$ RB.
\begin{center}
$\mathbf{H}(t,f)=\sum_{c=1}^{N_c}\mathbf{H}_c(t,f)$
$\mathbf{H}(t,f)=\sum_{c=1}^{N_c}\{\mathbf{P}_c+\mathbf{R}_c\circ{\mathbf{Q_c}(t,f)}\}$ 
\end{center}
where ``$\circ$'' is Hadamard product,\\
$N_c$ is number of clusters of scatterers (equivalently cluster of incidental rays impinging on BS with significant power),\\
subscript ${c}$ in $\mathbf{P}_c$ and $\mathbf{R}_c$, is cluster index,\\
$\mathbf{Q_c}(t,f)$ is temporal-frequential-variation controlling matrix having i.i.d. zero mean \& unit variance entries across matrix dimensions and certain correlation along time \& frequency dimensions.\\
$\mathbf{P}_c$ and $\mathbf{R}_c$ are spatial variation controlling matrices. $\mathbf{P}_c$ and $\mathbf{R}_c$ define the average spatial statistics of channel. $\mathbf{P}_c$ has complex entries representing the orientation of clusters and antenna-phase relationship. $\mathbf{R}_c$ has real \& non-negative entries representing angular spread of cluster i.e. the ratio of cluster distance to cluster size. Average signal power from $c^{th}$ cluster is jointly represented by $\mathbf{P}_c$ and $\mathbf{R}_c$.\\

\subsection{Underlying propagation phenomena}
To extend the effect of underlying propagation phenomena from single antenna system to MIMO (or massive MIMO), antenna-phase relation needs to be clarified. In single antenna system, there is no significant role of random phase because whatever cumulative random phase is there, it is tracked by PLL at receiver but in case of MIMO (or massive MIMO), random phases across antennas, also play role in creating uncorrelated channel vectors. Thus, there is a need to account two factors to model the channel for massive MIMO: one is that instead of considering Rician distribution (only for amplitude distribution), Complex non zero mean Gaussian distribution (which is the origin of Rician distribution) is to be considered. The second factor is that the deterministic antenna-phase relation must be considered in addition to the first factor because unfavorable propagation condition, in massive MIMO system, is some sort of drift from random to deterministic antenna-phase relationship. Determinism in antenna-phase relation is a function of angular spread of incoming radiation from a cluster. Slope of deterministic linear antenna-phase relationship depends on the direction of cluster relative to orientation of BS antenna array. If direction of a cluster makes an angle $\theta$ with BS antenna array axis then, the slope of deterministic antenna-phase relationship, for $\lambda/4$ antenna spacing, is given by $\frac{\pi}{2}{cos(\theta)}$ as shown in Figure 1. Physical interpretation of angular spread of incoming radiation is size or distance of the cluster. Here it is important to note that this relationship describes only deterministic part of antenna-phase relationship. Over all variation of phase across antennas is also contributed by random phases.
If instantaneous received power at $i^{th}$ BS antenna from $j^{th}$ MT via $c^{th}$ cluster, is $\beta_{ijc}$ then, it is simple to follow that total average power at $i^{th}$ BS antenna from $j^{th}$ MT is sum of the average received powers from all clusters i.e.

\begin{center}
$\beta_{ij}=\sum_{c=1}^{N_c}\mathbb{E}[\beta_{ijc}]$ where,
\end{center}
\begin{center}
$\beta_{ijc}=|p_{ijc}|^2+r_{ijc}^2$
\end{center}
Here channel coefficients are considered uncorrelated across the clusters. $p_{ijc}$ and $r_{ijc}$ are $(ij)^{th}$ elements of $\mathbf{P}_c$ and $\mathbf{R}_c$ respectively. $p_{ijc}$ and $r_{ijc}$ denote spatial mean and spatial standard deviation of $h_{ijc}$ respectively where $h_{ijc}$ is the channel gain for $(ij)^{th}$ link through $c^{th}$ cluster.

\subsection{Temporal-frequential-variation controlling matrix}
$\mathbf{Q_c}(t,f)$ is used to simulate temporal \& frequential variation and correlation. This type of channel matrix can be generated using its temporal or frequential (along RB index) covariance matrix. Since $q_{ijc}(t,f)$ is the $(ij)^{th}$ element of $\mathbf{Q_c}(t,f)$ thus, the matrix\\

$
\tilde{\mathbf{Q}}_{ijc}:=
\begin{bmatrix}
	q_{ijc}(1,1) 			& q_{ijc}(1,2) 			& ... & q_{ijc}(1,f_{mx})\\
	q_{ijc}(2,1) 			& q_{ijc}(2,2) 			& ... & q_{ijc}(2,f_{mx})\\
	....      				& ....      				& ... & ....\\
	q_{ijc}(t_{mx},1) & q_{ijc}(t_{mx},2) & ... & q_{ijc}(t_{mx},f_{mx})
\end{bmatrix}
$\\

is the set of channel coefficients in different RBs for communication link between $i^{th}$ BS antenna and $j^{th}$ MT via $c^{th}$ cluster when the channel coefficients are i.i.d. Gaussian (zero mean \& unit variance) distributed across dimension of channel matrix for a given RB and a given cluster. Column and row vectors of $\tilde{\mathbf{Q}}_{ijc}$ are jointly distributed. When temporal covariance matrix is considered, the column vectors of $\tilde{\mathbf{Q}}_{ijc}$ are distributed as: $CN(\mathbf{0},\mathbf{\Sigma_{col}})$ where $\mathbf{\Sigma_{col}}$ identifies the correlation among channel coefficients across RBs along time. Physically, $\mathbf{\Sigma_{col}}$ describes the Doppler spread of the channel. When frequential covariance matrix is considered, the row vectors of $\tilde{\mathbf{Q}}_{ijc}$ are distributed as: $CN(\mathbf{0}^T,\mathbf{\Sigma_{row}})$ where $\mathbf{\Sigma_{row}}$ identifies the correlation among channel coefficients across RBs along frequency. Physically, $\mathbf{\Sigma_{row}}$ describes the delay spread of the channel.

\subsection{Spatial variation controlling matrices}
$\mathbf{P}_c$ and $\mathbf{R}_c$ describe spatial variation of channel or equivalently, variation in channel across antennas. $\mathbf{P}_c$ and $\mathbf{R}_c$ control the spatial mean and spatial variance of $\mathbf{H}(t)$ respectively. 
Following description is given regarding the method to generate spatial variation controlling matrices:
\begin{itemize}
\item A given cluster subtends an angle on BS antenna array depending on its size and distance from receiver thus, incoming radiation from a cluster has angular spread approximately equal to the subtended angle. Incoming radiation of limited angular spread, is equivalent to a significant component in certain direction which can be modeled by non-zero mean complex Gaussian distribution (Rician distribution in amplitude) with a variance and mean depending on angular spread, direction of cluster and average received power from cluster.
\item Value of $r_{ijc}$ (element of $\mathbf{R}_c$) is proportional to angular spread of incoming radiation at $i^{th}$ BS antenna from $j^{th}$ MT via $c^{th}$ cluster. $r_{ijc}=\beta_{ijc}$ for $2\pi$ $radian$ angular spread and $r_{ijc}=0$ for zero $radian$ angular spread.
\item For each cluster, the required complex mean vector (columns of $\mathbf{R}_c$) for $j^{th}$ column of $\mathbf{H}(t)$ is generated by generating phase vector and amplitude vector separately.
\item Phase vector for given cluster, is generated by selecting the entries of vector from modulo $2\pi$ linear antenna-phase relation with randomly selected slope (relating to direction of cluster) as shown in Figure 1 (for $\lambda/4$ antenna spacing). Similar antenna-phase relation for line of sight radiation is presented in \cite[sec 4.2]{Ngo2014}.
\item Amplitude vector for a given cluster, is calculated element-wise, by the relation: $|p_{ijc}|^2+r_{ijc}^2=\beta_{ijc}$. Parameters $p_{ijc}$, $r_{ijc}$, $\beta_{ijc}$, $\beta_{ij}$ and normalization factor for channel matrix are inter-related and they follow different profiles along $i$, $j$ and $c$ indexes depending on environment, but it is not covered in current work.
\end{itemize}

\begin{figure}
\psscalebox{1 1} 
{
\begin{pspicture}(-0.5,-1.837353)(6.631311,1.837353)
\psline[linecolor=black, linewidth=0.04, arrowsize=0.05291667cm 3.0,arrowlength=1.4,arrowinset=0.0]{<-}(0.9923527,1.8426471)(0.9923527,-1.0373529)
\psline[linecolor=black, linewidth=0.04, arrowsize=0.05291667cm 3.0,arrowlength=1.4,arrowinset=0.0]{<-}(6.7523527,-0.7173529)(0.8323527,-0.7173529)
\psline[linecolor=black, linewidth=0.04](2.2723527,-0.7173529)(2.2723527,-0.87735295)
\psline[linecolor=black, linewidth=0.04](3.5523527,-0.7173529)(3.5523527,-0.87735295)
\psline[linecolor=black, linewidth=0.04](4.8323526,-0.7173529)(4.8323526,-0.87735295)
\psline[linecolor=black, linewidth=0.04](6.112353,-0.7173529)(6.112353,-0.87735295)
\psline[linecolor=black, linewidth=0.04](0.9923527,0.082647055)(0.8323527,0.082647055)(0.8323527,0.082647055)
\psline[linecolor=black, linewidth=0.04](0.9923527,0.88264704)(0.8323527,0.88264704)
\psline[linecolor=black, linewidth=0.04, linestyle=dotted, dotsep=0.10583334cm](3.5523527,-0.7173529)(3.5523527,0.88264704)
\psline[linecolor=black, linewidth=0.04, linestyle=dotted, dotsep=0.10583334cm](6.112353,-0.7173529)(6.112353,0.88264704)
\psline[linecolor=black, linewidth=0.04](0.9923527,-0.7173529)(3.5523527,0.88264704)
\psline[linecolor=black, linewidth=0.04](3.5523527,-0.7173529)(6.112353,0.88264704)
\rput[bl](0.5123527,0.082647055){$\pi$}
\rput[bl](0.3523527,0.88264704){$2\pi$}
\rput[bl](2.1123526,-1.357353){4}
\rput[bl](3.3923528,-1.357353){8}
\rput[bl](4.672353,-1.357353){12}
\rput[bl](5.9523525,-1.357353){16}
\psdots[linecolor=black, dotsize=0.04](6.4323525,-1.357353)
\rput[bl](2.1123526,-1.837353){Antenna position}
\rput{31.419533}(0.36916128,-2.1354156){\rput[bl](3.980588,-0.4114706){Slope=$\frac{\pi}{2}cos(\theta)$}}
\rput[bl](1.2135292,1.5273529){$\theta$: Direction of shrinkage relative}
\rput[bl](1.740588,1.1602942){to antenna array axis}
\rput{-270.071}(-0.2618546,-0.88076186){\rput[bl](0.30999976,-0.57147056){Relative phase}}
\psdots[linecolor=black, dotsize=0.04](6.7523527,-1.357353)
\psdots[linecolor=black, dotsize=0.04](6.592353,-1.357353)
\end{pspicture}
}\\
{Fig. 1: Antenna-phase relation in massive MIMO under unfavorable propagation condition (deterministic part of channel)}
\end{figure}

\subsection{Effect of clusters}
The proposed channel model translates clustering of scatterers into clustering of channel gains which can be observed in histogram plot in Figure 2. For a given cluster, clustering of channel gains is proportional to the ratio of cluster distance to cluster size (exact or approximate relation has not been established yet). Over all clustering of channel gains is stronger when the number of clusters is small and there is a cluster with dominant power (received at BS) as well as having larger ratio of cluster distance to cluster size. The stronger the clustering of channel gains, the stronger the determinism in channel. Unfavorable propagation is a subset of determinism in channel where dominant scatterers of two or more MT-BS links are clustered in such directions that clusters make approximately equal angle with BS antenna array axis (here equal angle is in magnitude sense e.g. clusters having physical directions $30^o$ and $150^o$ relative to BS antenna array axis, have equal angle in magnitude sense).

\section{Insights on simulated channel}
Simulation is performed for single and multiple clusters per MT with constant total power ($\beta_{ij}=1$) and with varying power profile. Number of BS antennas ($N$) is taken 128 and 20. Elements of $G=\frac{1}{N}H^HH$ are denoted by $g^{nd}_{ij}$ for $i\neq j$ (non-diagonal) and by $g^{d}_{ij}$ for $i=j$ (diagonal). The metric $g^{nd}_{ij}$ is a measure of cross-correlation among MTs which is key parameters for different decoders and precoders proposed for massive MIMO.

\subsection{Histogram of channel coefficients}
Histogram of elements of a complex channel vector for a given MT, given cluster and given time instant, is plotted in Figure 2 and Figure 3 with $r_{ijc}=0.1$ and $r_{ijc}=0.5$ respectively. As described in channel model, the $r_{ijc}$ is the spatial variance of channel coefficient ($h_{ijc}$). The smaller the variance, the higher the determinism in the channel along spatial dimensions. As the spatial determinism increases in the channel (because of reducing angular spread), the histogram of channel coefficients converges to a ring on complex plane. For $r_{ijc}=\beta_{ijc}$, elements of channel vector become i.i.d. Gaussian random variables. For $r_{ijc}=0$, elements of channel vector become completely deterministic lying on a circle on complex plane with linearly varying phase along BS array. The histogram of channel coefficients resembles with measured channel in \cite{Larsson2014}.
\begin{figure}
\psscalebox{0.78 0.78} 
{
\begin{pspicture}(-0.5,-4.2)(11.2,4.2)
\rput(5.1,0.0){\includegraphics[scale=0.74]{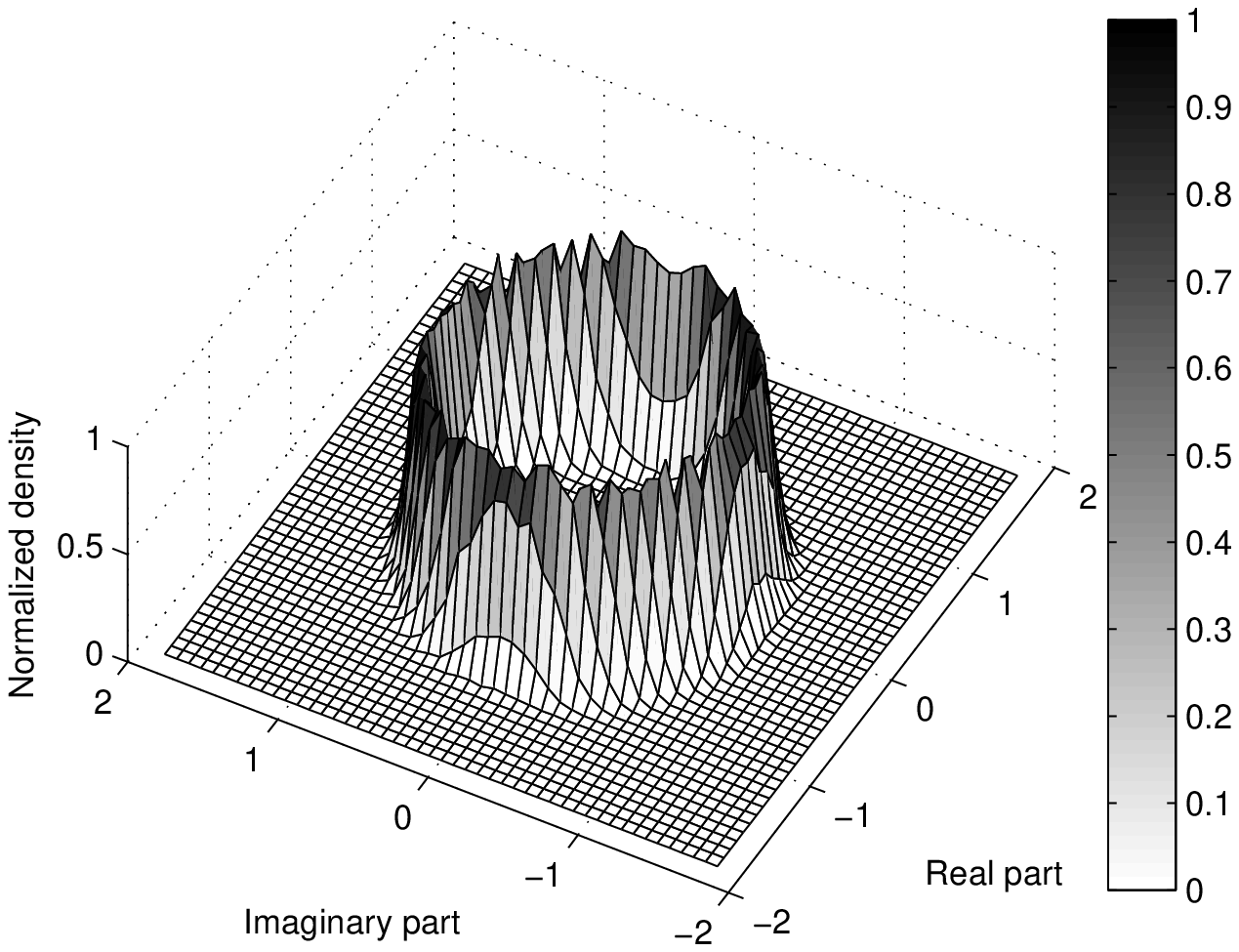}}
\end{pspicture}
}
Fig 2. Histogram of elements of channel vector with $r_{ijc}=0.1$ (small angular spread or nearly LOS)
\end{figure}
\begin{figure}
\psscalebox{0.78 0.78} 
{
\begin{pspicture}(-0.5,-4.2)(11.2,4.2)
\rput(5.1,0.0){\includegraphics[scale=0.74]{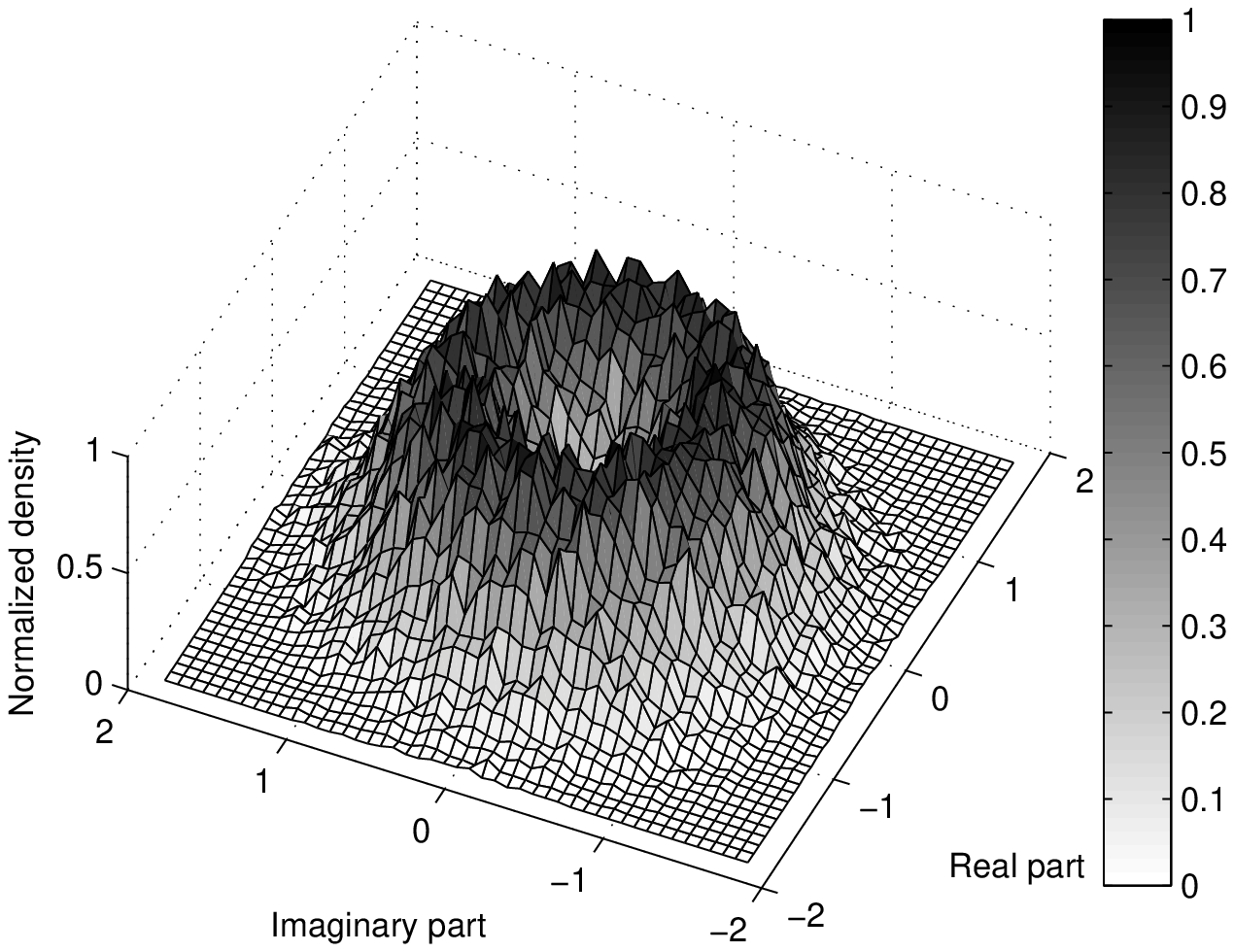}}
\end{pspicture}
}
Fig 3. Histogram of elements of channel vector with $r_{ijc}=0.5$ (large angular spread or nearly NLOS)
\end{figure}\\

\subsection{Histogram of cross-correlation in MTs}
Since the value of $|g^{nd}_{ij}|$ depends both on the angular separation and directions of clusters so this part of simulation presents the average statistics of channel for randomly selected directions of two clusters, in terms of histogram of $|g^{nd}_{ij}|$. For i.i.d. Gaussian case ($r_{ijc}=\beta_{ijc}$ i.e. $r_{ijc}=1$), there is an insignificant probability that $|g^{nd}_{ij}|$ is greater than a certain value ( 0.4 in this simulation) while in case of higher degree of determinism ($r_{ijc}=0.1$), $|g^{nd}_{ij}|$ can take value up to 1 with a significant probability. However, there is an advantage of higher degree of determinism that histogram plot shifts towards zero which can improve the average performance of the system.

\begin{figure}
\psscalebox{0.78 0.78} 
{
\begin{pspicture}(-0.5,-4.2)(11.2,4.2)
\rput(5.1,0.0){\includegraphics[scale=0.74]{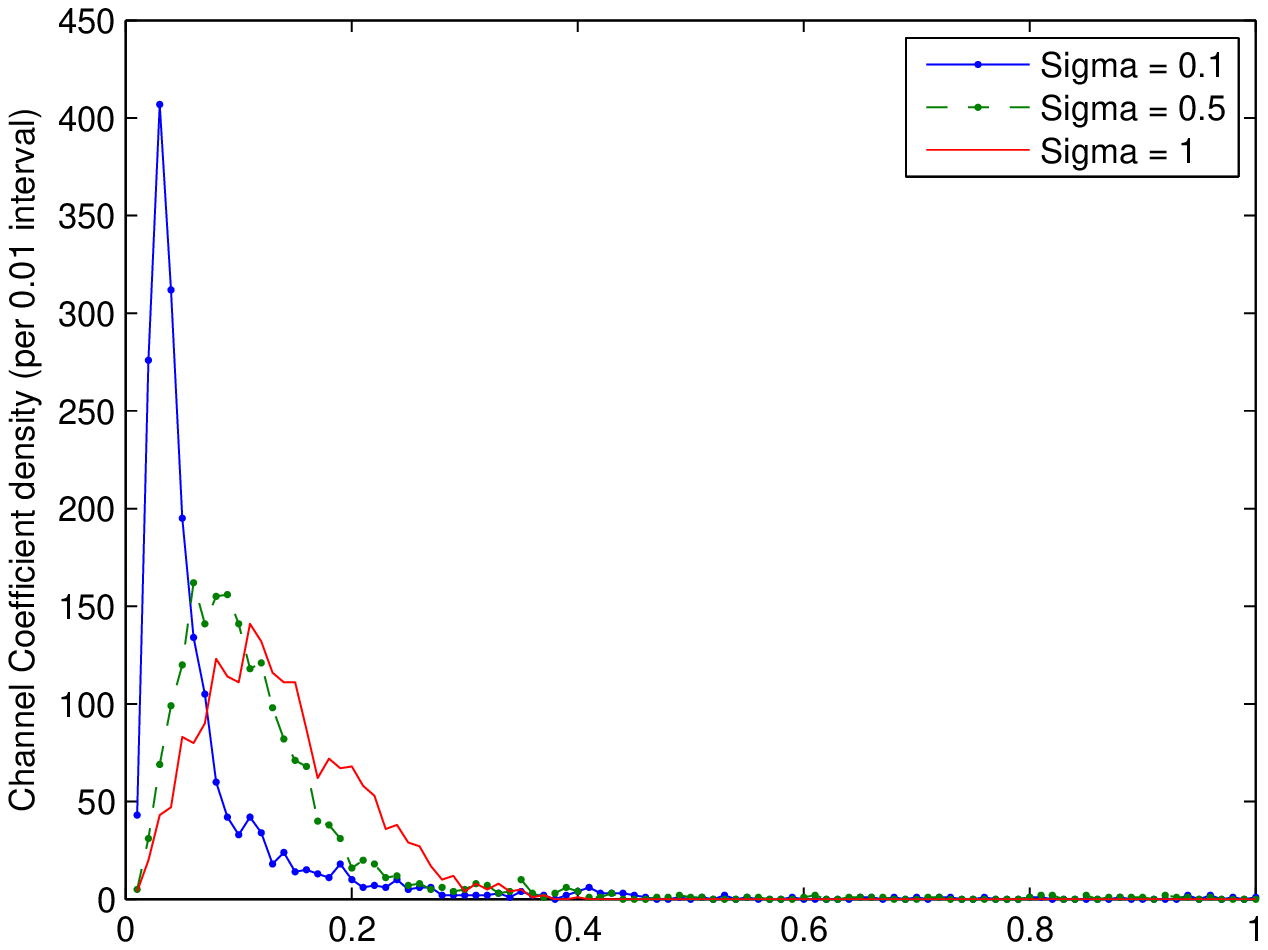}}
\rput{0}(-0.8636478,-2.6542475){\rput[bl](2.96,-1.4){Cross-correlation of MTs ($|g^{nd}_{ij}|$)  $\rightarrow$}}
\rput{0}(-0.8636478,-2.6542475){\rput[bl](7.96,4){Sigma $=r_{ijc}$}}
\end{pspicture}
}
Fig 4. Histogram of Cross correlation in channel vectors for different degrees of determinism (2000 total samples).
\end{figure}

\section{Validation of channel model}
To compare simulated channel with measured channel parameters of \cite{Payami2012}, three clusters of scatterers with different power and different angular spread are used with each of 6 MTs. Channel is simulated for i.i.d. ($r_{ijc}=\beta_{ijc}$) , NLOS ($r_{ijc}$ much greater than zero) and line of sight (LOS) i.e. $r_{ijc}$ close to zero. 

\subsection{User correlation}
User correlation is related to inner product of channel vectors. User correlation is estimated from simulated channel with 1000 random realizations for 3 LOS and 3 NLOS MTs. Correlation matrix A corresponds to 3 NLOS MTs and 128 BS antennas, B corresponds to 3 LOS MTs and 128 BS antennas, C corresponds to 3 NLOS MTs and 20 BS antennas, D corresponds to 3 LOS MTs and 20 BS antennas. Result is well matching with measured data \cite[Sec III(F)]{Payami2012}. Number of BS antennas has direct impact on cross-correlation of MTs. This channel model provides additional insight on relatively smaller cross-correlation in LOS environments. Increasing spatial determinism in the channel, as described in previous section, shifts the histogram of cross-correlation towards zero.\\ 
  
$A:
\begin{bmatrix}
1.000  &  0.299  &  0.177\\
0.299  &  1.000  &  0.085\\
0.177  &  0.085  &  1.000\\
\end{bmatrix}
$
$
B:
\begin{bmatrix}
1.000  &  0.062  &  0.044\\
0.062  &  1.000  &  0.056\\
0.044  &  0.056  &  1.000\\
\end{bmatrix}\\
$\\

$
C:
\begin{bmatrix}
1.000  &  0.361  &  0.300\\
0.361  &  1.000  &  0.339\\
0.300  &  0.339  &  1.000\\
\end{bmatrix}
$
$
D:
\begin{bmatrix}
1.000  &  0.215  &  0.173\\
0.215  &  1.000  &  0.202\\
0.173  &  0.202  &  1.000\\
\end{bmatrix}
$\\

\subsection{Distribution of eigenvalues}
Cumulative distribution function (CDF) of eigenvalues of $\frac{1}{N}H^HH$ is plotted in Figure 5, using 1000 random realization for LOS \& NLOS MTs and 128 \& 20 BS antennas. This result also resembles with measured data \cite[Sec III(D)]{Payami2012}. However, distribution is plotted for normalized average power unlike the \cite[Sec III(D)]{Payami2012} where actual received power is used. Effect of smaller number of BS antennas on spreading of eigenvalues, can be seen in red curve (NLOS with N=20). Additional insight on eigenvalue distribution is that eigenvalues confines to unity with increasing spatial determinism because the fluctuation in power along space reduces with increasing spatial determinism.\\

\begin{figure}
\psscalebox{0.75 0.75} 
{
\begin{pspicture}(-0.5,-4.2)(11.2,4.2)
\rput(5.1,0.0){\includegraphics[scale=0.74]{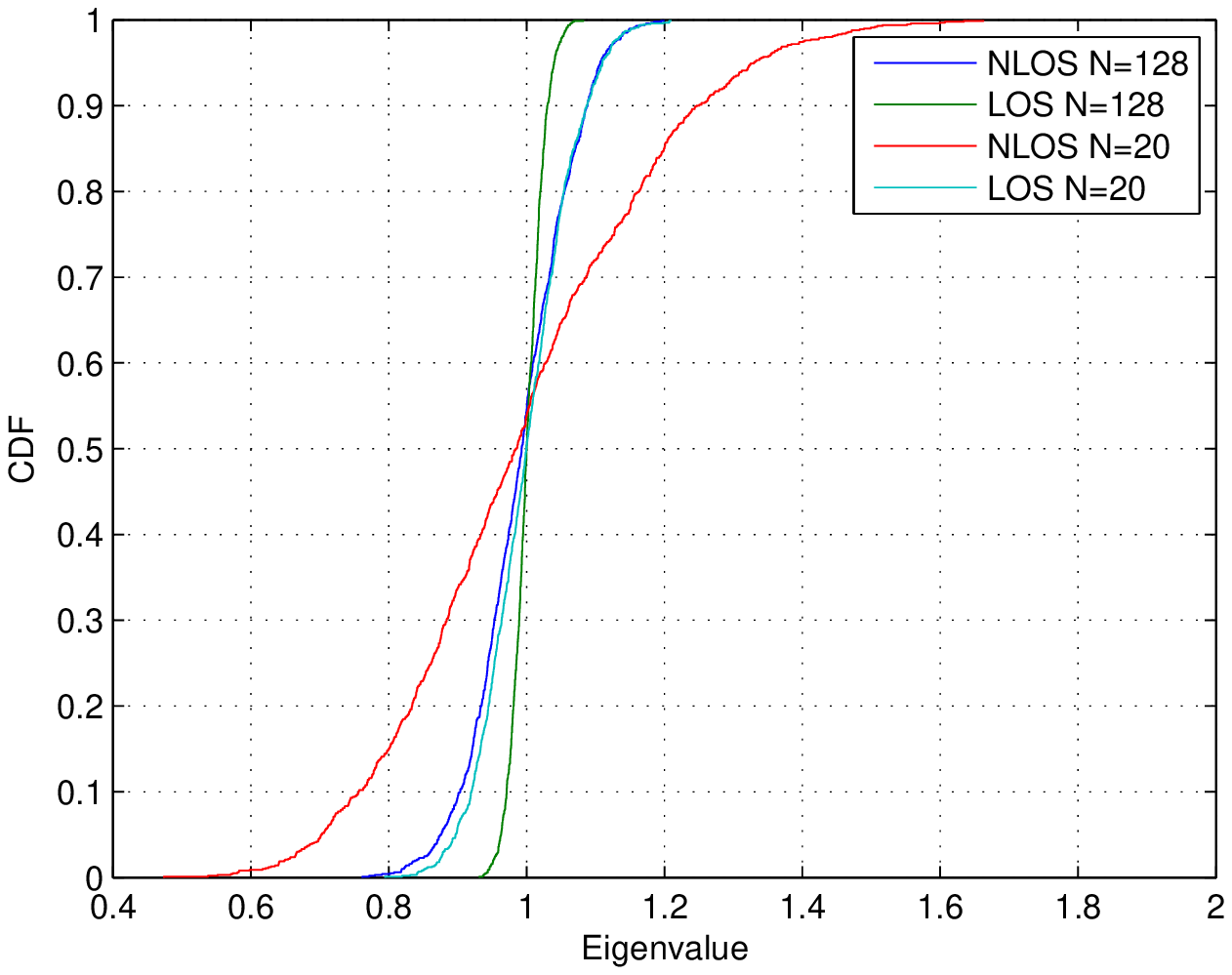}}
\end{pspicture}
}
Fig 5. Distribution of eigenvalues of $\frac{1}{N}H^HH$
\end{figure}

\subsection{Distribution of power over antenna array}
Average received power in $(ij)^{th}$ BS-MT link, is $\beta_{ij}$. This part of simulation considers the varying $\beta_{ij}$ with antenna index $i$. Appearance-disappearance of scatterers along array is assumed to be basic cause of large scale fading across BS antennas in existing literature but this work put an another insight on this part. Average received power from a cluster can vary with BS antenna index even if there no shadowing at all. For example, the orientation of some of the dominant reflectors in the cluster can change the average received power with antenna index. Thus, the simulation of large scale fading using superficial geometry of clusters, may not give precise results. Modeling of large scale fading in this work, is done directly by average received power per cluster. Average received total power along BS array in Figure 6, resembles with measured channel parameters \cite[Sec III(C)]{Payami2012}.

\begin{figure}
\psscalebox{0.75 0.75} 
{
\begin{pspicture}(-0.5,-4.2)(11.2,4.2)
\rput(5.1,0.0){\includegraphics[scale=0.74]{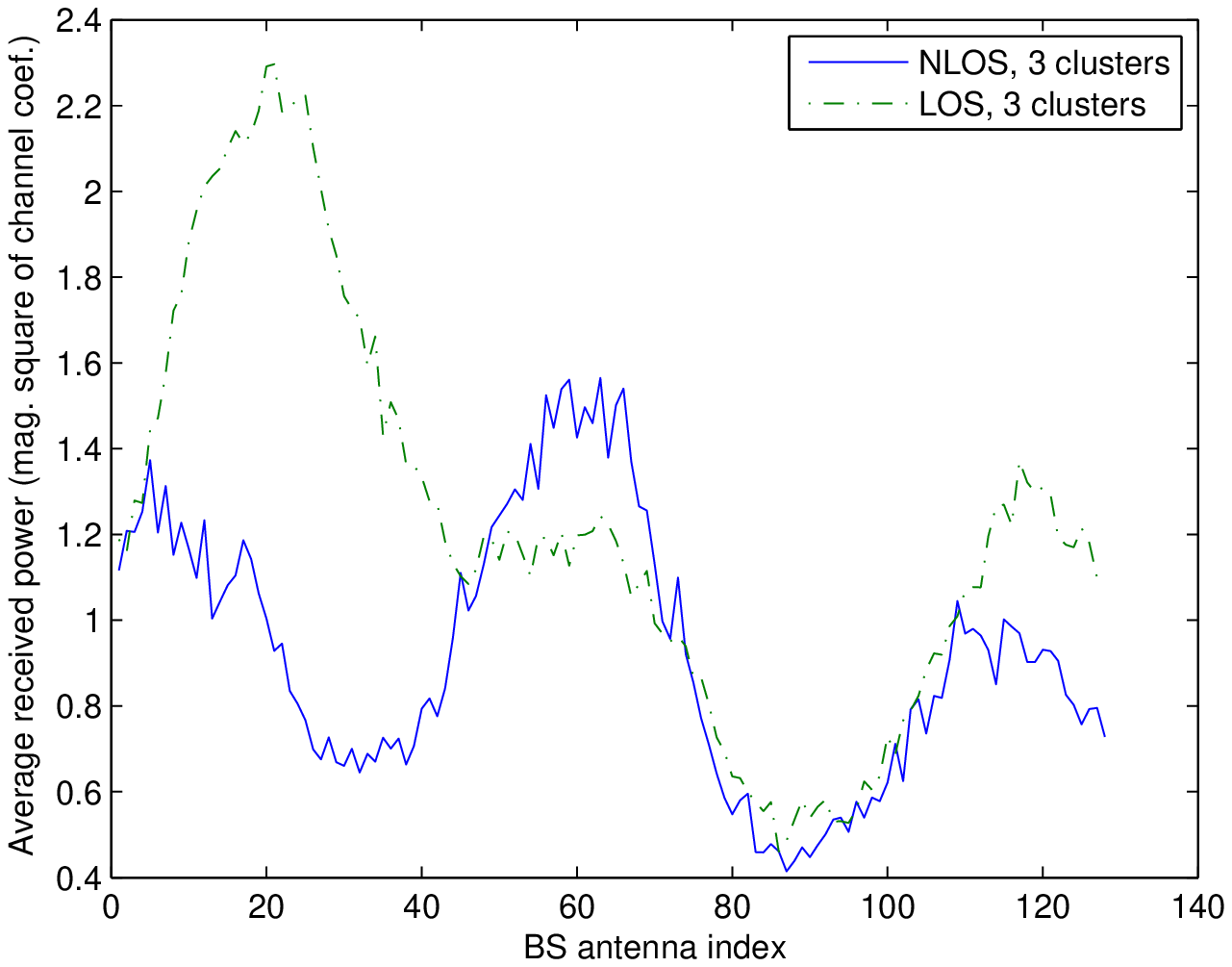}}
\end{pspicture}
}
Fig 6. Average power distribution along BS antenna array
\end{figure}

\section{Conclusion}
A statistical block fading channel model for multiuser massive MIMO is presented in this paper. The presented channel model incorporates the properties of two types of channel model: CBSCM and GBSCM. This channel model enables the statistical simulation of multidimensional arrays so that temporal or frequential and spatial correlation in channel can be simulated simultaneously. Temporal, frequential \& spatial variations and underlying propagation phenomena, are explained in conjunction with different parameters of channel model. Insights on spatial determinism in channel is presented revealing the hidden advantages of spatial determinism in channel for massive MIMO. The parameters of simulated channel model, are identified to resemble with existing measured channel parameters for massive MIMO. Future scope for this channel model is the generalization of model i.e. simulation for more than two dimensions simultaneously e.g. time, frequency and space.

\renewcommand{\baselinestretch}{1}


\end{document}